\documentclass[preprint,11pt]{elsarticle}
\usepackage{xspace}
\usepackage{multirow}
\usepackage{ifthen}
\usepackage{xcolor,colortbl}
\usepackage{tabulary}

\newboolean{authnotes}
\setboolean{authnotes}{false}

\ifthenelse{\boolean{authnotes}}
{
\newcommand{\amy}[1]{\footnote{{\bf Amy: #1}}}
\newcommand{\usman}[1]{\footnote{{\bf Usman: #1}}}
\newcommand{\valerio}[1]{\footnote{{\bf Valerio: #1}}}
\newcommand{\ale}[1]{\footnote{{\bf Alessandro: #1}}}
\usepackage{draftwatermark}
\SetWatermarkLightness{0.85}
\SetWatermarkScale{5}
\newcommand{\whereis}[1]{}
\newcommand{\memo}[1]{{\color{red} #1}}

}
{
\newcommand{\amy}[1]{}
\newcommand{\usman}[1]{}
\newcommand{\valerio}[1]{}
\newcommand{\ale}[1]{}
\newcommand{\whereis}[1]{}
\newcommand{\memo}[1]{#1}
\usepackage{draftwatermark}
\SetWatermarkText{}
}
\newcommand{\dbp}{\textsc{DBP}\xspace}
\newcommand{\dbs}{\textsc{MBS}\xspace}
\newcommand{\fakeparagraph}[1]{\vspace{1mm}\noindent\textbf{#1}.}

\newcommand{\pbdc}{prediction-based data collection\xspace}
\newcommand{\Pbdc}{Prediction-based data collection\xspace}

\newcommand{\sleep}{\textit{sleep}\xspace}
\newcommand{\aactive}{\textit{active}\xspace}
\newcommand{\hibernation}{\textit{hibernation}\xspace}
\newcommand{\standby}{\textit{standby}\xspace}

\newcommand{\indoor}{\textsc{intel}\xspace}
\newcommand{\tunnel}{\textsc{tunnel}\xspace}

\makeatletter
\define@key{Gin}{Trim}
           {\let\Gin@viewport@code\Gin@trim\expandafter\Gread@parse@vp#1 \\}
\makeatother

\usepackage[tight,footnotesize]{subfigure}

\usepackage{balance}

\journal{Ad Hoc Networks}

\begin{document}

\begin{frontmatter}
%
\author[unitn]{Usman Raza\corref{cor1}}
\ead{usman.raza@toshiba-trel.com}

\author[uniurb]{Alessandro Bogliolo}
\ead{alessandro.bogliolo@uniurb.it}

\author[uniurb]{Valerio Freschi}
\ead{valerio.freschi@uniurb.it}

\author[uniurb]{Emanuele Lattanzi}
\ead{emanuele.lattanzi@uniurb.it}

\author[fbk]{Amy L. Murphy}
\ead{murphy@fbk.eu}

\cortext[cor1]{Corresponding author}

\address[unitn]{Toshiba Research Europe Limited, Bristol, BS1 4ND, United Kingdom}

\address[uniurb]{University of Urbino, Piazza della Repubblica 13, Urbino, Italy}

\address[fbk]{Bruno Kessler Foundation, Via Sommarive 18, Trento, Italy}

\title{A Two-Prong Approach to Energy-Efficient WSNs: Wake-Up Receivers plus Dedicated, Model-Based Sensing \tnoteref{t2}}
\tnotetext[t1]{A preliminary version of this paper appeared in the proceedings of $9^{th}$ IEEE International Symposium on Industrial Embedded Systems (SIES 2014)~\cite{sies14} }

\address{}
\sloppy

\begin{abstract}
Energy neutral operation of WSNs can be achieved by exploiting the idleness of the workload to bring the average power consumption of each node below the harvesting power available.
This paper proposes a combination of state-of-the-art low-power design techniques to minimize the local and global impact of the two main activities of each node: sampling and communication. Dynamic power management is adopted to exploit low-power modes during idle periods, while asynchronous wake-up and prediction-based data collection are used to opportunistically activate hardware components and network nodes only when they are strictly required. Furthermore, the concept of ``model-based sensing'' is introduced to push prediction-based data collection techniques as close as possible to the sensing elements. The results achieved on representative real-world WSN case studies show that the combined benefits of the design techniques adopted is more than linear, providing an overall power reduction of more than 3 orders of magnitude.

\end{abstract}


\end{frontmatter}


\section{Introduction}
\label{sec:intro}

Advances in energy harvesting (EH) have begun to shift the decade-old research
goal of the WSN community from energy efficiency to energy
autonomy. Nevertheless, building a self-sustaining EH-WSN remains a
challenging task, especially in indoor environments where the harvested energy
can be several orders of magnitude less than the consumed energy. For example,
a typical, optimized application can spend milliwatts in sensing and
communication, while the typical harvesting range for an indoor harvester
remains on the order of microwatts~\cite{zhang13, tirnity}.  Therefore, work
to reduce power consumption is still required to achieve energy autonomy.

Dynamic power management (DPM) represents the cornerstone to any power
reduction approach. By targeting workload idleness and putting a node's
micro-controller (MCU) into a low-power inactive mode, DPM saves energy.  In
such a state, consumption drops from hundreds of milliwatts, seen for high
consuming activities such as radio transmission, to a fraction of a
microwatt. Clearly as nodes spend more time in this low-power state, system
lifetime improves. Nevertheless, the activities of a typical node, including
sampling, channel listening and receiving, and transmitting, require the MCU
to switch to high power modes, with each transition additionally incurring a
cost. Our work proposes a unique combination of hardware and software 
techniques to reduce the frequency and duration of various
high-consumption tasks, not only increasing the time a node can spend in the
low-power state, but decreasing the number of transitions out of this state. This 
results in huge energy savings without compromising the application requirements.


\memo{We propose a novel technique for WSN applications with periodic data
collection, such as those where sensor data forms the input of a control loop,
thus requiring timely communication of the sensed values at a central
controller.} Our proposed architecture employs (1) a wake-up receiver
alongside (2) a dedicated sensing peripheral that incorporates (3) a data
reduction software module. The first and third techniques aim to decrease the
cost of data collection by focusing on the use of the radio, as it represents
one of the most power hungry components on the node, while the second
technique achieves additional savings by reducing the cost of frequent sensor
sampling. The combination of these techniques results in very long idle
periods with rare occurrences of power hungry tasks such as radio
transmissions and receptions.  Low-power modes, provided by multiple
underlying hardware components, are exploited by the dynamic power manager
during the idle periods to conserve energy.

To illustrate the potential of these techniques, we briefly address each,
beginning with the communication.  Traditional system architectures employ
sophisticated duty cycling medium access control (MAC) techniques
to achieve significant reductions in consumption by putting the radio to sleep
for extended periods. Nevertheless, nodes still waste energy in two primary
ways.  First, they must periodically wake up and listen to the channel in case
a node is attempting to transmit to it.  If there is nothing to receive, this
energy spent listening is wasted, leading to idle listening.  On the other
side, a sender with data to communicate must transmit until the receiver wakes
up, often leading to long transmission times among unsynchronized nodes.
Wake-up receivers are a novel hardware approach to eliminate these two main
sources of overhead. Specifically, they provide an ultra low power receiver
that is always on and listening to the channel, either the same channel used
for communication or a dedicated, out-of-band channel. When a packet is to be
transmitted, a preamble is generated by the transmitter to trigger the wake-up
of the data radio on the receiving node.  This eliminates idle listening
by turning on the main radio module only when there is a packet to be
received.  Further, it reduces the transmission time by ensuring that the
receiver is ready to receive immediately after transmission of the preamble,
thus avoiding the repeated transmissions typical of duty cycling protocols.

While exploiting a wake-up receiver leads to significant lifetime gains in
many scenarios, applications that collect data at high frequency still incur
significant costs to transmit the raw data.  To ameliorate these costs,
several techniques have been proposed to reduce the amount of data sent
without compromising the application requirements. In the
technique we consider, each node calculates a model for its data and
communicates this model to the sink. The sink then uses these models to
predict the data samples at each node. As long as the real samples closely
match the model, no data is communicated, however as soon as the real data
deviates significantly from the data estimated by the model, a node generates
a new model and transmits it to the sink. Such approaches have the potential
to eliminate 90 to 99\% of the transmissions~\cite{tkde}, depending on the
type of data being sampled and the sophistication of the model.

Interestingly, the combination of a data reduction technique and the wake-up
receiver reduces the radio cost to a point where periodic sensor sampling,
normally considered a low-consumption task, actually accounts for a
significant portion of the energy consumption. Therefore, to make the sampling
cheaper, we introduce a novel hardware-software based technique called
Model-based Sensing (\dbs). \dbs delegates the tasks of periodically sampling
the sensors and the running data reduction algorithm to a dedicated ultra-low
power hardware peripheral. This peripheral operates without involving the
power hungry primary MCU as long as the data reduction technique is successful
in suppressing the sensed data. Only rarely, when there is a need to transmit
data, is the main MCU turned on to transmit the data.

By combining these techniques, we increase the length of the idle periods of
the MCU and the radio transceiver, which each offer multiple low-power modes
to save energy. Additionally, ancillary components such as flash memories and
real-time clocks, can be opportunistically turned on and off to save power.
In general, the low-power modes and component states must be carefully managed
to meet the workload and minimize consumption. This paper shows the potential
savings of this combination of techniques in concrete, real-world case studies
representative of many periodic data collection applications. The resulting
system consumes only a few microwatts, resulting in a 
system lifetime improvement of \emph{three orders of magnitude}, and reaching
the point where indoor energy harvesters can sustain node operation.

In summary, the contributions of this paper include:
\begin{itemize}
\item a system architecture that uniquely combines three software and hardware
  techniques to drastically reduce node workload and offload the frequent and
  expensive activities from the main MCU. The \dbs module is presented for the
  first time in this paper.  Our combination results in ultra-lightweight
  applications with long idle periods between consecutive activities.

\item a case-study based evaluation of the savings achievable when the
  low-power modes of the radio, MCU and various peripherals exploit the
  increased idleness achieved by our architecture. To fully examine the
  contributions of each technique, we evaluate them both individually as well
  as in various combinations.





\end{itemize}

The rest of the paper is organized as follows: Section \ref{sec:relwork} provides an overview of related work on data reduction techniques and ultra low-power WSNs; Section \ref{sec:sysarch} describes the overall system architecture; Section \ref{sec:mbs} introduces the design principles of model-based sensing; Section \ref{sec:setup} and Section \ref{sec:results} describe the experimental setup and present the results respectively; Section \ref{sec:conclusions} concludes the work.


\section{Background and Related Work}\label{sec:relwork}

This section offers background on the techniques used throughout this paper to
effectively exploit DPM: prediction-based data collection, wake-up
receivers and ultra low-power sensor nodes.


\subsection{\Pbdc} 
\label{subsec:pbdc}

As a first step toward reducing consumption, we turn to techniques that reduce
the data that must be transmitted. Many such techniques exist such as data
compression, in-network data processing and data
aggregation~\cite{anastasi2009energy}, but we focus on \pbdc due to its
simplicity and demonstrated effectiveness on \memo{scenarios with periodic data
collection, e.g., where a WSN serves as a component in a control loop.}

In \pbdc, the original application-required sampling period is maintained, but
the total amount of data transmitted is reduced~\cite{anastasi2009energy} by
generating a model for the sensed data. This model is used at the sink to
approximate the sampled data points.  With each new sample, the node verifies
that it falls within allowable error tolerances.  If so, no action is taken,
but if not, a new model is generated and transmitted to the sink.  If the
model closely approximates the data trend, the network communication is
significantly reduced, up to 99\% in some
cases~\cite{raza12:what,edgemining,pwsn}. Various types of models have been
studied.  Probabilistic
models~\cite{DBLP:conf/vldb/DeshpandeGMHH04,DBLP:conf/icde/ChuDHH06}
approximate data with a user-specified confidence, but special data
characteristics must be encoded by domain experts.  Alternate techniques
employ linear regression~\cite{DBLP:conf/ewsn/TuloneM06}, autoregressive
models~\cite{DBLP:conf/mswim/TuloneM06} and Kalman
filters~\cite{Jain:2004:ASR:1007568.1007573}, but with sizeable memory and
computational requirements, making them difficult to implement on
resource-limited motes. A simpler, linear approach~\cite{raza12:what},
detailed in Section~\ref{sec:dbp}, was recently proposed by some of the
authors of this paper, and is adopted for the case study here.


\subsection{Wake-up Receiver} 
Wake-up receivers are a viable solution to enable low-power, asynchronous
communication, essentially by triggering the activation of the primary
radio~\cite{lin04}. Several wake-up solutions have recently been
developed~\cite{hambeck11,huang11,marinkovic11,petrioli14,magno14,magnotii14},
each optimizing different parameters such as working power, sensitivity,
distance range, latency, and operating frequency. These solutions use the
radio channel to convey triggering signals, but alternative proposals use
out-of-band signals such as ultrasound. Notably, this paper evaluates two
wake-up modules for the VirtualSense platform i) an ultrasonic wake-up receiver recently developed by
some of the authors of this paper~\cite{lattanziUS13} and ii) a radio wake-up receiver presented in~\cite{magno14} . Both outperforms
state-of-the-art radio wake-up receivers with a sub-$\mu$A quiescent current
consumption.

Recent researches have shown the potential for asynchronous wake-up schemes
in some application scenarios~\cite{jelicic12,sies14}. Specifically,
studies~\cite{tn,petrioli} have shown that, in comparison to duty
cycling protocols, wake-up receivers offer longer system lifetimes, lower
latency, and better reliability in multiple application domains. These studies
note that wake-up receivers are particularly effective for applications with
ultra-low traffic, as they avoid the idle listening incurred at routing nodes
to periodically check for incoming packets.  This paper builds on this
observation, specifically showing how traditional applications with moderate
data rates can be converted into ultra-low traffic applications with \pbdc,
thus enhancing the effectiveness of wake-up receivers and DPM.

\subsection{Ultra-low Power Platforms}
Careful design of embedded wireless hardware platforms is key to energy
efficient WSNs. Over the last 15 years, many efforts have been devoted to
devise novel ultra-low power MCUs and radio transceivers. \memo{First generation WSN platforms included
  Rene~\cite{hill00}, Mica~\cite{hill02} and
  Telos~\cite{polastre05}. Afterwards, the design of embedded architectures
  for sensor nodes has mainly relied on 16-bit MCUs and IEEE 802.15.4
  compliant transceivers. Recent trends show a shift towards 32-bit
platforms that meet strict WSN energy requirements~\cite{andersen14}. All of
these platforms come with the ability to opportunistically turn off hardware
components in periods of idleness. As the runtime power consumption depends on various dynamic power management features available within these 
motes, this paper focuses on one single mote named VirtualSense,  a
platform designed by some of the authors of this paper.} The next section focuses on the
various elements VirtualSense offers for dynamic power management.


\section{System Architecture}
\label{sec:sysarch}

\begin{figure}[t]
\centering
\includegraphics[width=.95\columnwidth]{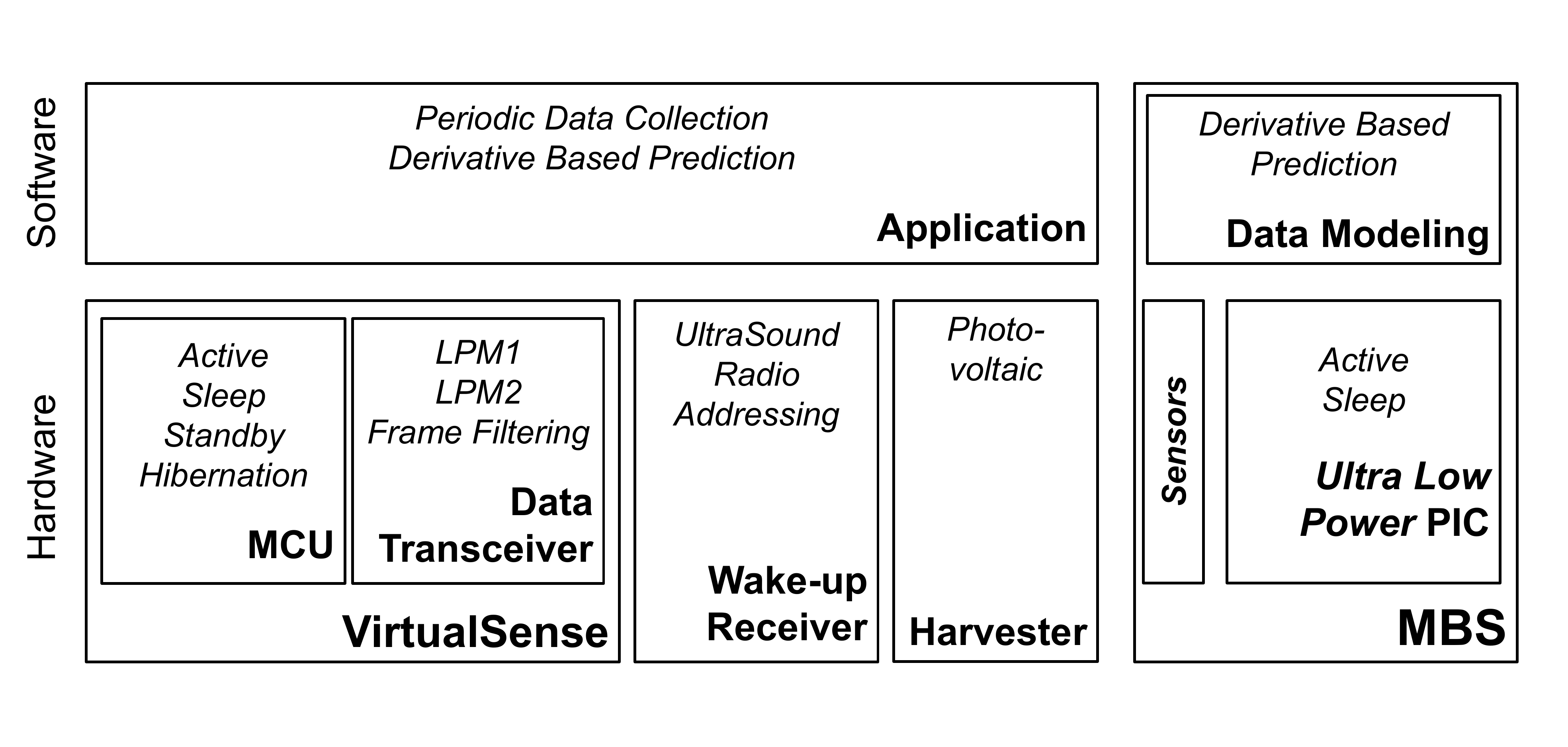}
\caption{High level system architecture showing the primary configurable
  components and the configuration options available for each of them.}
\label{fig:arch}
\end{figure}

The primary contributions of this paper are first, a novel combination of
technologies from hardware to software to achieve fully energy autonomous
systems and second, the concrete, case-study based evaluation of multiple
configurations of these technologies.  Figure~\ref{fig:arch} offers a very
high level overview of the configurable components we consider, dividing them
between software and hardware, then further dividing
the hardware among those belonging to the
VirtualSense~\cite{lattanzi_virtualsense_12} platform (the MCU and the data
transceiver), the wake-up receiver, the model based sensing (MBS) element
(\dbs), and the energy harvester.  This section offers a high level
description of most of these components, emphasizing their configuration
options. As MBS itself is a novel contribution of this paper, we discuss it
separately in Section~\ref{sec:mbs}.

\subsection{Derivative-Based Prediction (DBP)}
\label{sec:dbp}

To reduce the amount of data transmitted by each node, we developed an easily
implementable data prediction technique that captures the data trends. With
DBP, first described in~\cite{raza12:what}, we adopted a linear model computed
from $m$ data samples, the first and last $l$ points we refer to as edge
points.  The linear model is calculated as the slope of the line connecting
the average of the first $l$ edge points with the average of the last $l$ edge
points.  This computation resembles the calculation of the derivative, hence
the name \emph{Derivative-Based Prediction}.

On initialization, $m$ points are collected, then the first model is generated
and sent to the sink.  Subsequently, each sensor sample is checked against the
value the model predicts.  If the reading is within a given tolerance, no
action is taken as the sink will also use the model to approximate the sensor
sample. However, if the application tolerances are exceeded, a new model is
generated and transmitted to the sink.

To offer a brief example, consider an outdoor light sensor.  At sunrise the
linear, \dbp model will be an upward sloping line.  At some point, however,
the light levels will cease to increase and the upward sloping model will be
replaced with a flatter model, corresponding to the daytime light
levels. While this explanation is over-simplified, it offers the intuition of
\dbp.

It is worth mentioning that we have applied \dbp to several real data sets
ranging from soil temperature to indoor temperature and light
values~\cite{tkde}.  In all cases, a reasonably tuned \dbp produces data
reduction rates above 89\% and in most cases above 98\%. These savings are
sufficient to exploit the combination of technologies explored in this paper.

\dbp, or any data reduction approach, can either be run on the main MCU or on
pushed closer to the hardware and run as part of \dbs sensing peripheral. We
discuss the latter case in Section~\ref{sec:mbs} and highlight its benefits in
Section~\ref{sec:results}.

\subsection{VirtualSense}
\label{sec:vs}
VirtualSense~\cite{lattanzi_virtualsense_12} is an open-hardware ultra low-power sensor node featuring a Java-compatible virtual runtime environment. The software stack of VirtualSense is based on the Contiki operating system (OS)~\cite{dunkels04} and the Darjeeling Virtual Machine (VM)~\cite{brouwers09}, suitably modified to make it possible for a Java programmer to fully exploit the low-power states of the underlying MCU, a MSP430F5418a as well as the data transceiver, a CC2520.

\subsubsection{Microcontroller Unit}
VirtualSense features four categories of power states: \aactive, \standby, \sleep, and \hibernation. In \standby  the CPU is not powered, but the clock system is running and the unit is able to wake itself by means of timer interrupts. In \sleep both the CPU and the clock system are turned off and the unit is woken  up only by an external interrupt. In \hibernation even the memory system is switched off and there is no data retention requiring a complete reboot of the OS at wake-up, together with a restore of the VM heap. 

Power consumption varies significantly across different states of VirtualSense. 
In the \aactive mode, the average power consumption is approximately 13mW when processing and 66mW for transmitting, while the consumption reduces to 14.67$\mu$W in \standby, 1.32$\mu$W in \sleep, and 0.36$\mu$W in \hibernation. We also note that the time to transition from one state to another is non-negligible. Specifically, the transition to \aactive is 25ms from \standby and \sleep, and 500ms from \hibernation. 
WSN applications that do not need to maintain execution state can also use a
memory-less hibernation mode ({\it ML Hibernation}), which does not preserve
the VM state, reducing the wake-up time to 27ms. In summary, with effective
DPM, power savings of several orders of magnitude can be achieved.

%

\subsubsection{Data Transceiver} VirtualSense provides several low-power
communication features by exploiting the inactive modes (LPM1 and LPM2) of the
CC2520 radio transceiver, as well as the hardware frame filtering (FF)
capabilities that prevent the reception of non-intended
packets. 

The CC2520 is controlled by the main MCU through the serial peripheral
interface bus (SPI) and six general purpose I/O pins (GPIOs). LPM2 is the
lowest power consumption mode in which the digital voltage regulator is turned
off, no clocks are running, no data is retained, and all analog modules are in
the power down state. In this state,  power consumption is around 13.5$\mu$W,
but the embedded controller needs to be rebooted at wake-up. In LPM1, the
digital voltage regulator is on, but no clocks are running. In this state,
power consumption is around 3mW, while all data/configurations are retained and the analog modules can be controlled by the main MCU.
During transmission, the power consumption of the entire transceiver, due to the RF module plus the embedded microcontroller, ranges from 48.6mW (at -18dBm output power) to 100.8mW (at +5dBm output power), while in standard receive mode the power consumption is 69.9mW.

The frame filtering function (FF) rejects non-intended frames not matching the local address. With FF enabled, the transceiver can be switched immediately to the LPM2 in case of a non-intended frame without a need to process the rest of the frame and wake up the MCU~\cite{lattanzi14}. 

\subsection{Wake-up Receivers} Table \ref{tab:wurscomparison} reports the
primary features of the two wake-up receivers(WuRs) used in this study,
characterized by their use of ultrasound or radio waves.

\begin{table*}
\begin{center}
\caption{Main parameters of ultrasonic and radio wake-up modules.}
\begin{footnotesize}
\begin{tabular}{|l|c|c|}\hline
{\bf WuR type} & {\bf US} &{\bf Radio} \\ \hline\hline
{\bf Radiation pattern} & $55^\circ$at -6dB & Omnidirectional \\ \hline
{\bf Frequency} & 40KHz & 868MHz \\ \hline
{\bf Range} & 15m & 20m \\ \hline
{\bf Tx power Standby} & 40nW & 690nW \\ \hline
{\bf Tx power Active} & 37mW & 78mW \\ \hline
{\bf Rx power Listening} & 1640nW & 462nW \\ \hline
{\bf Rx power Decoding} & 14$\mu$W & 49$\mu$W \\ \hline
{\bf Throughput} & 20bps & 10Kbps \\ \hline
\end{tabular}
\end{footnotesize}
\vspace{-1mm}
\label{tab:wurscomparison}
\end{center}
\end{table*}

\subsubsection{Ultrasonic WuR}
The ultrasonic WuR considered in this work consists of a transmitter and of a
receiver based on piezoelectric transducers working at 40KHz with a 2KHz
bandwidth~\cite{lattanziUS13}. Nodes equipped with this WuR are triggered upon
detection of an ultrasonic carrier signal. Optional selective triggering can
be obtained by encoding 8~bit addresses through \textit{On-Off-Key} (OOK)
modulation of the carrier.

The receiver consumes 1640nW during listening periods and 14$\mu$W when decoding the preamble or the address. Power consumption by the transmitter is 40nW in standby and 37mW in active mode used to encode and transmit the wake-up signal. These parameters are compatible with a 15m operating range. The overall throughput of the system amounts to 20bps. Wake-up triggering times range from 50ms in non-addressing mode to 450ms  for 8-bit addressing mode. Ultrasound transducers are notably directional in that the main radiation lobe is 55$^\circ$ at -6dB.

\subsubsection{Radio WuR}
The radio wake-up solution we adopted is based on a recently developed ultra
low power radio WuR~\cite{magno14}. This module works at 868MHz within a 20m
range; it is equipped with a high sensitivity receiver (-42dBm) with a power
consumption of only 462nW (while listening) and of 49$\mu$W (during
decoding). The power attributable to the transmitter is only 690nW in standby
and 78mW during transmission. Similar to the ultrasonic WuR, this radio WuR
features the capability to selectively address nodes, thanks to a mechanism
based on OOK modulation. Wake-up times range from 130$\mu$s without addressing
to 0.8ms for selective triggering with a 1~byte address (more than two orders
of magnitude lower than that of the ultrasonic WuR) and the overall throughput
is 10Kbps. The radiation pattern of the associated antenna is omnidirectional.

\begin{table}[tb]
\begin{center}
\caption{Amount of harvestable energy available in indoor
  environments. \cite{Basagni2013, calhoun2005design}}
\begin{footnotesize}
\setlength\extrarowheight{2.5pt}
\begin{tabular}{|l|c|}\hline
{\bf Harvester} & {\bf Power density} \\ \hline\hline
Photovoltaic &  less than 10 $\mu$watt/cm$^2$ \\ \hline
Electromagnetic & 1-4  $\mu$watt/cm$^3$\\ \hline
Vibration (electrostatic) &  3.8 $\mu$watt/cm$^2$\\ \hline
Radio Frequency &  0.1 $\mu$watt/cm$^2$ \\ \hline
Acoustic noise & 0.003-0.096 $\mu$watt/cm$^3$\\ \hline
\end{tabular}
\end{footnotesize}
\label{tab:indoorharvest}
\end{center}
\end{table}

\subsection{Energy Harvester}
\label{sec:harvester}
In this paper, our objective is to reduce node consumption to a rate that can
be supported in a typical indoor environment with state of the art energy
harvesting techniques.  Table~\ref{tab:indoorharvest} outlines five such
technologies, indicating the amount of harvestable energy each provides. These
values are orders of magnitude lower than those in typical outdoor
environments.

The WSNs deployed in the both case studies considered in this paper are exposed
to low-intensity artificial light from lamps. This light can be readily
converted by a photovoltaic cell into electrical energy. Therefore,
photovoltaic cells optimized for low illuminance are an ideal choice. In this
paper, we consider the Panasonic AM-1816~\cite{am1816}, a palm-sized cell
designed to support small electronics indoors, even under low-intensity
fluorescent lights.



\section{Model-based Sensing}
\label{sec:mbs}

Traditionally, WSN nodes are designed to periodically \emph{sense} the
environment and then immediately \emph{send} this data to a sink node. However,
several applications interrupt this procedure, supressing the transmission of
many sensed data, either through prediction-based data collection techniques
as briefly discussed in Section~\ref{subsec:pbdc} or transmitting only
significant events such as a detected forest fire, volcanic eruption, or
landslide. The commonality between prediction-based data collection and
event-driven applications is that both sense the data periodically, then
process it to determine whether or not to send any data. When effective, these
techniques avoid transmission in a large majority of the cases, nevertheless,
the mote's power hungry MCU must still be woken up to sense and process the
sample. Turning on the MCU frequently incurs a significant cost for both
switching to and staying in a high power mode. Our observations show that this
sensing cost accounts for the most significant energy consumption once the
communication costs are drastically reduced with a data reduction technique
and a wake-up radio. To reduce this sensing cost, we introduce a novel
\emph{hardware-software} technique called \emph{model-based sensing (MBS)},
which moves the sensing and processing tasks from relatively power hungry MCU
to a dedicated hardware peripheral.


%
%
%
%
%
%


Figure~\ref{fig:arch} shows our approach with the MBS module containing
sensors and and ultra-low power microcontroller on the hardware side and a
data processing logic on the software side. This software can either implement
the functionality of an event-driven application or a data reduction technique
such as \pbdc. In this paper we focus on the latter, noting that MBS proposes
\textit{moving} the data modeling software from the primary MCU (labeled
``application'' in the figure) into the MBS module, allowing sampling and
processing without requiring the expensive, primary MCU. Only when samples
fall outside the model must the MCU be turned on, together with the
primary radio transceiver, for transmission of a new model to the sink. We
select the microcontroller for MBS module such that it consumes orders of
magnitude less energy than main MCU, enabling significant energy savings.

%

\fakeparagraph{Implementation Details} In this paper, we use Derivative-based
Prediction~\cite{raza12:what} as the basis for \dbs. The model is piece-wise
linear and is computed based on $m$ samples. Hence, each sample needs to be
digitized, stored in memory, and processed in order to decide whether a new
model must be transmitted or not. Processing entails integer sums, divisions,
and comparisons that can be performed by an off-the-shelf ultra low power
peripheral interface controller (PIC).  The reference architecture proposed in
this paper is based on Microchip PIC16F1825, which draws only 36nW in sleep
mode with data retention and 10.8$\mu$W while running at 32KHz. It
accommodates 12~16-bit ADC channels and 1024~bytes of RAM, which can hold up
to 512~ADC samples. 
We chose this PIC because its
power consumption in processing and sleep modes is three orders of magnitude
less than that of the MCU used in our VirtualSense mote. 
 
Although, in principle, pure hardware implementations could be envisioned to minimize the number of components to be embedded into the sensing layer, the cost and the power consumption of state-of-the-art PICs do not motivate the development of application-specific integrated circuits.

\section{Experimental Setup}
\label{sec:setup}

Our unique combination of the technologies presented in
Section~\ref{sec:sysarch} and Section~\ref{sec:mbs} exploits the lowest
consumption modes of the VirtualSense platform for as much time as possible,
moving to a low-power MCU state between activities such as transmissions and
receptions. Using \dbp decreases the costs in multiple ways. Consider that in
any system, a node transmits its own data, forwards data from other nodes, and
unnecessarily overhears packets destined to other nodes. Each of these events
requires the node and the primary data radio to be switched to a high power
consumption mode.  \dbp reduces the total traffic in the network, thus
reducing the frequency of all these events, and consequently increasing the
time the node can remain in the lowest power mode. However, with \dbp running
as the main application on the mode, the main MCU still needs to process each
periodic sample, resulting in relatively high sampling costs. We have solved
this problem by delegating the tasks of sampling and running \dbp to a
specialized ultra-low power sensing peripheral called \dbs, obviating the need
for expensive and frequent wake-ups of main MCU. Moreover, overheads related
to idle listening and message overhearing are further reduced by using a
wake-up receiver and the hardware frame filtering capabilities of the radio
transceiver. Our multifaceted hardware software approach intends to radically
reduce the power consumption of the WSNs.

The results presented in the next section come from a series of simulations
with \dbp, VirtualSense, wake-up receivers and \dbs, performed with actual
data collected from two real-world WSNs deployed in a road tunnel and an
office building. The estimated power consumption is based on real, empirical
measurements of power consumed by VirtualSense and the wake-up receivers in
different power modes. This section offers details on both case studies, the
power measurements and concludes with an estimation of energy harvested from
photovoltaic cells, as our final goal is an energetically sustainable system.

\subsection{Case Studies}

The two case studies described in this section correspond to a real-world
deployment and a testbed, representative of sparse and dense WSNs
respectively. Both applications require periodic data samples to be shared
with a central control system.

\subsubsection{\tunnel: Adaptive Lighting in Road Tunnels}

\begin{figure*}[t]
\centering
\includegraphics[page=2, width=\textwidth]{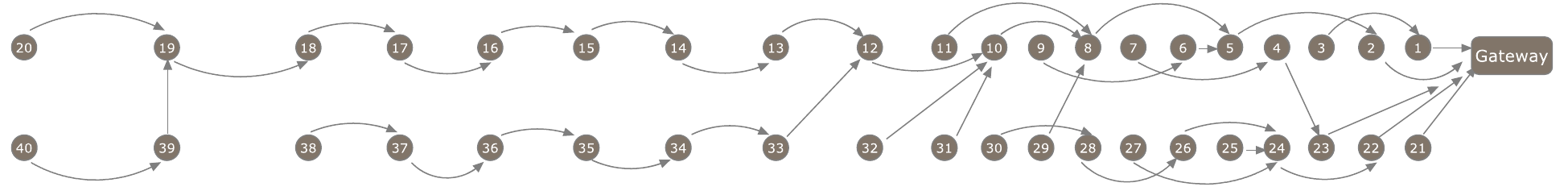}
\vspace{-5mm}
\caption{Layout of nodes in the tunnel and a sample data collection tree with
  a depth of 15 hops.}
\vspace{-4mm}
\label{fig:topology_tunnel}
\end{figure*}

Our first case study is based on a pilot deployment in a real road tunnel in
Trento, Italy~\cite{ceriotti11:tunnel}.  In this 260~m tunnel, a WSN of 40 nodes is deployed to periodically measure the light
levels.  The nodes close to the entrance are exposed to sunlight, while nodes
deep in the tunnel only detect the artificial light from the lighting system.
In all cases, the light levels detected by the sensors every 30~s are
transmitted over a 15-hop network, shown in Figure~\ref{fig:topology_tunnel},
to reach a gateway at the entrance of the tunnel. The values are then used by
a control system to gradually adjust the intensity of the lamps throughout the
tunnel to meet the legislated light levels. The control system was designed in
collaboration with lighting engineers to tolerate a limited amount of data
loss and to accommodate some degree of error in the quality of the sensed
values.

In this paper, we evaluate the system's power consumption with multiple
hardware and software configurations. While the next section addresses the
hardware configurations, here we consider the application layer, as it is
affected by the case study itself.  Specifically, we must consider the amount
of data reported by each node with and without the \dbp data prediction
algorithm. For this, we used actual data traces collected from the tunnel over
a 47-day winter period, a total of $5,414,400$ samples.  Without \dbp, we
assume each sample is transmitted by each node immediately after being
sampled.  With \dbp, instead, each node transmits only when the model
changes. When running \dbp only on the \dbs module, we assume that the main
MCU and data transceiver would not be turned on unless there is a model
change. For the purposes of this study, we configured \dbp to allow the
predicted data values to deviate from the actual light values by at most 5\%
or 15~lux.  Further, at most two consecutive samples can fall outside this
error bound before a new model generation is triggered.  These settings allow
\dbp to reduce the total sent traffic by 99.74\% w.r.t. periodic
reporting~\cite{tkde}.


\memo{The data collection topology, shown in Figure~\ref{fig:topology_tunnel}, is generated using an implementation of unit disk graph model  in Cooja simulator~\cite{cooja}. The communication range is set to 15m, the minimum range of the two WuRs in Table 1, while the interference range is set to double of the communication range. This topology along with the total network traffic is fed to the power consumption model described in Section~\ref{sec:simulation} to estimate the power consumption.}

\subsubsection{\indoor: Indoor Environmental Monitoring}
Our second study is based on arguably one of the first publicly available
datasets collected from a WSN, specifically an indoor, 54~node Mica2Dot
deployment inside the Intel Berkeley Research Lab~\cite{buonadonna05, madden05}. The laboratory is approximately $40\times31~m^2$.  The dataset consists of 36 days worth of
environmental data sampled every 31~s including light, temperature and
humidity, resulting in 2.3~million values for each sensor type.

As with the tunnel, this data displays daily cycles, with light showing the
most irregularities as the humans in the environment directly affected it. We
configured \dbp such that the predicted data values can deviate from the
actual values by at most 5\% or 15~lux for light, 1\% for humidity and 0.5
degree Celsius for temperature. This configuration reduced the periodic transmissions by 97.58\% for light, 99.50\% for humidity and 99.60\% for temperature. 

\begin{figure}
\centering
\includegraphics[width=.95\columnwidth]{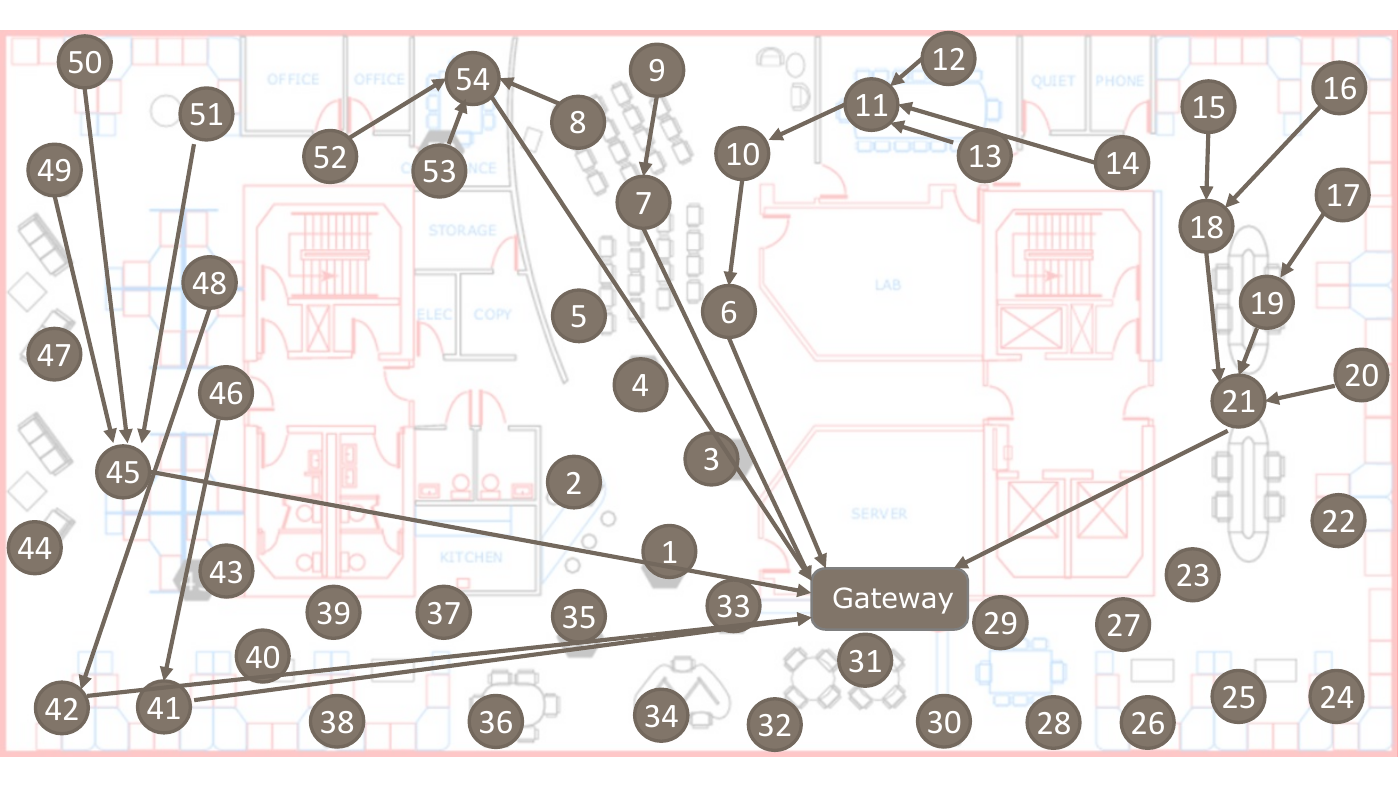}
\vspace{-3mm}
\caption{Layout of nodes in the Intel-lab testbed together with the data collection tree.
 For the sake of readability, direct connections between nodes not involved on multi-hop paths are not represented.}
\vspace{-4mm}
\label{fig:topology_intel}
\end{figure}

\memo{To model the real-world channel and interference conditions experienced by the main data transceiver, we use publicly available aggregate connectivity data~\cite{madden05}. Based on the provided link qualities, we construct a 4-hop collection tree, as shown in Figure~\ref{fig:topology_intel}. Given these link qualities and the total amount of traffic sent and forwarded over the collection tree, number of link level transmissions has been estimated and fed to our power consumption model described next. }

\subsection{Power Models and Simulations}
\label{sec:simulation}
\begin{figure}[t]
\centering
\includegraphics[width=7cm]{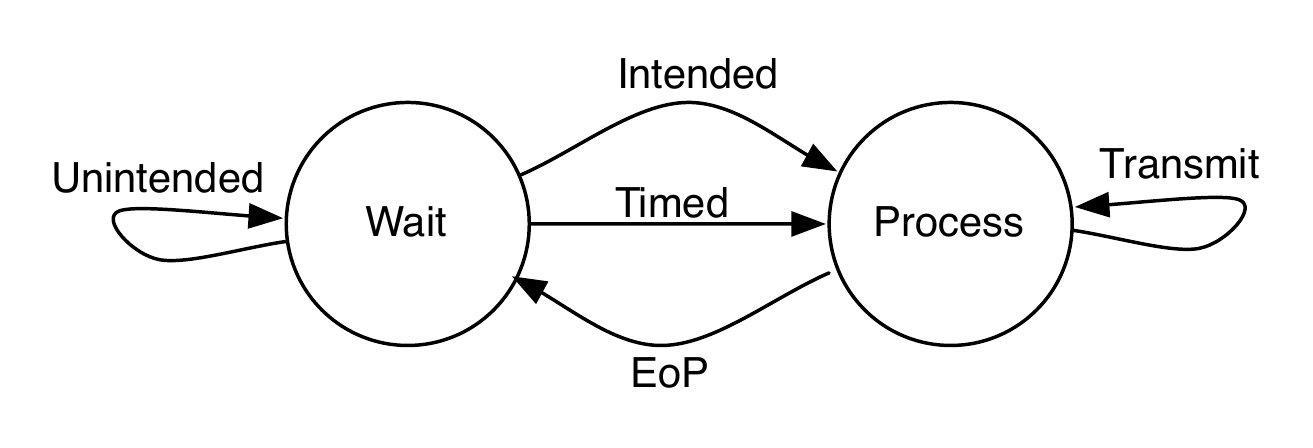} 
\caption{\memo{Reference functional state diagram.}}
\label{fig:PSM}
\end{figure}

Power simulations were obtained starting from the functional state diagram reported in Figure~\ref{fig:PSM}, which describes the behavior of a WSN node capable of exploiting idle periods during its workload in order to save power. 

\memo{The state diagram is represented in terms of states and transitions. In particular, the \textit{Wait} state represents a family of inactive modes fully accessible and exploitable by the dynamic power manager. In \textit{Wait} state, the node reacts to three types of events: {\it i)} reception of an intended packet; {\it ii)} overhearing of an unintended packet;  {\it iii)} wake up from timed interrupt for periodic tasks (i.e., sampling). The \textit{Process} state is representative of activities such as sampling a given physical quantity and evaluating the need to transmit a sample according to the prediction strategy, or relaying an incoming packet toward destination (i.e the sink). When the processing ends (EoP), the node transitions back to the \textit{Idle} state.

The functional state diagram allows us to derive the following five operating conditions accounting for node power consumption: 
\begin{enumerate}
\item
waiting; 
\item 
hearing an unintended packet;
\item
receiving and routing a packet;
\item
waking up autonomously to sample then transmit;
\item
waking up autonomously to sample but not transmit thanks to \dbp.
\end{enumerate}

The power consumption model was used to accurately estimate the consumption based on the runtime behavior of VirtualSense nodes \cite{bogliolo13}. In particular, we computed the average power consumption as a weighted average, where weights are represented by the actual occurrence rates of each operating condition as derived from real-world traffic data (for each of the two case studies).
}
Indeed, the datasets obtained from the case studies were fed to the simulator, which captures the behavior and power consumption of each node, taking into account
all the configuration options of the hardware and software components involved, as shown in Figure~\ref{fig:arch}. In details, the simulator allows us to: 
\begin{itemize}
\item
select the low-power mode of the MCU;
\item
select the low-power modes of the data radio transceiver (with and without hardware frame filtering);
\item
choose between the ultrasonic or radio wake-up modules;
\item
activate either software DBP, or MBS.  
\end{itemize}
When the wake-up receiver is not in use, we assume the primary radio runs the standard ContikiMAC protocol with a 100ms wake-up interval.

\memo{
The power consumption values of the MCU \cite{bogliolo13}, the radio transceiver \cite{bogliolo13}, the ultrasonic \cite{lattanziUS13} and the radio wake-up \cite{magno14} modules, and the new MBS module were accurately characterized by means of real-world measurements. 

It is worth noticing that the simulator takes traffic statistics from real \indoor testbed or from \tunnel simulation in Cooja~\cite{cooja}. Hence, the radio and acoustic channels are not modeled within the simulator. Nevertheless, the effects of the data radio channel are inherently represented by the input data according to the models or experimental conditions used to generate them, as detailed in Section 5.1. 

As for the radio/acoustic wake-up communication, all the experiments were conducted within the nominal operating conditions of the devices (in terms of range and orientation), where the effects of non-idealities are negligible. Within these conditions, the triggering reliability is near-perfect. Pushing the wake-up technologies beyond their nominal conditions would require a thorough analysis of the effects of-false positives (false-negative) wake-up events on power consumption (data packet loss). Such analysis goes beyond the scope of this work.
}

\subsection{Estimation of Harvestable Energy}
To estimate the harvestable energy for our nodes, we use the actual light
values collected in our case studies. \memo{These light values are already expressed in lux for \indoor. The raw sensor values from \tunnel sensors is also converted to lux, thanks to an accurate calibration process described in \cite{ceriotti11:tunnel}}. For estimating the harvestable energy, we assume that the same sensed illuminance is produced from a fluorescent light source and is incident on on a Panasonic AM-1816, a palm-sized photovoltaic cell, which then is
attached to a harvester described in ~\cite{porcarelli2012multi} with a charging efficiency of 79\%.
\memo{Our model for harvestable energy is then derived  by a piecewise linear model using  empirical measurements available for the photovolatic cell in~\cite{am1816} and the harvester in~\cite{porcarelli2012multi}.}


\section{System Evaluation}
\label{sec:results}


Our goal is to evaluate system power consumption with multiple different
configurations of the hardware and software components of our architecture, as
outlined in Figure~\ref{fig:arch}. This section reports the achievable savings
in both case studies, first discussing those enabled by various
combinations of these components, then highlighting those achieved by the
two WuRs and the novel \dbs module.

\subsection{Energy Savings}
For our case studies, Tables~\ref{tab:results1} and~\ref{tab:results2} show
$23$ different configurations, each of which corresponds to a specific
hardware states. The first row shows a standard node configuration that uses
the \textit{standby} mode of MCU and the \textit{LPM1} mode of the data
transceiver with no WuR. We consider this hardware configuration,
in combination with a software layer that does not use the \dbp prediction
scheme as the baseline.  Power consumption values, shown in $\mu$W, are
computed as averages over all the nodes of the WSN being studied.

When the data is sent periodically, without \dbp or \dbs,
energy efficiency is achieved by using a suitable low power hardware
configuration. We observed a lifetime improvement of 19x and 92x
(ID~11) for \tunnel and \indoor respectively. The larger
improvement for \indoor compared to \tunnel is due to the higher node density
of the former, which causes excessive overhearing and unnecessary triggering
of WuRs in most hardware configurations, resulting in higher
baseline consumption. The FF capabilities and addressing mode of the WuRs
(Radioa and USa) are,therefore, more effective in mitigating this effect for
the more dense \indoor deployment, enabling larger savings.
 
In all configurations, adding \dbp reduces the network traffic, reducing power
consumption. Additionally, as expected, increasing the use of low-power modes
also reduces consumption.  Without the WuRs (IDs~1-5),
power consumption reductions are modest, approximately eight times for both
case-studies, even when considering a configuration that exploits the \sleep
mode of the MCU and avoids unnecessary overhearing with the data radio (FF).

Intuitively, the addition of the WuR should have a significant
impact by reducing idle listening in the \tunnel and \indoor applications
where data is generated roughly only twice per minute. However, the observed
lifetime improvement varies significantly across the wake-up technologies. For
example, in \tunnel, the improvement is 19x for Radio (ID~11)
compared to only 3.3x for US (ID~7). A better performance of Radio
WuR is due to its three orders of magnitude better throughput and
therefore a lower wake-up triggering cost w.r.t. the US WuR.


Instead, combining the radio WuR with the extremely low data rate
of \dbp (ID~11, \dbp) results in a significant 463 times
improvement, a result that is much larger than the improvements attainable by
each technique in isolation.
This remarkable result is a concrete demonstration of the benefits of
eliminating idle listening with the very low transmission rates achieved with
\dbp. Additionally, as the node no longer needs to forward frequent data
packets on behalf of other nodes, it can spend more time between samples in
the low power mode of the MCU. Without \dbs, the sampling interval (30~s for
\tunnel and 31~s for \indoor) defines the maximum time a MCU can stay in a low
power mode. \dbs, instead, gets rid of this limitation by offloading the
sampling task from main MCU, enabling even longer idle periods for the
MCU. Maximally, with \dbs, the MCU can stay in low power mode between two
consecutive model updates. To offer an example, if a \tunnel node does not
observe significant variations in light throughout the night, a model will not
change and the MCU can likely stay in a low power mode for this whole
duration. These very long periods of idleness result in \emph{three orders of
  magnitude} lifetime improvement. Specifically, the addition of \dbs improves
the lifetime from 463x to \emph{2052x} for \tunnel (ID~11).

Missing entries in Tables~\ref{tab:results1} and~\ref{tab:results2} indicate
infeasible hardware configuration. For example, the hibernation mode of the
MCU has a shut-down and a wake-up time the cannot be accommodated with the
sampling rates of the target applications when \pbdc is not exploited to
reduce the traffic. Similarly, memory-less hibernation modes are not
compatible with the software implementation of \dbp, which requires data
retention. It is worth noting that all configurations become feasible with
\dbs, which achieves benefits by reducing the traffic, reducing the number of
MCU state transitions, and providing an auxiliary memory in the sensing layer
that enables ML hibernation.

In summary, these numerical results bolster our argument that the individual
techniques of prediction-based data collection, DPM, \dbs, and wake-up
receivers, while individually capable of achieving improvements, are even more
powerful when combined into a single system. The synergistic effect results in
more than 3 orders of magnitude improvement in lifetime in both case studies
presented here. The performance of the two WuRs and \dbs under
different hardware-software configurations is discussed in more detail in the
following subsections.

\begin{table*}[htb]
\begin{center}
\caption{System-wide energy savings in the tunnel case study. The gray cells
  indicate the baseline for calculating the power consumption improvement ratio of all
  other configurations. (Hib.=Hibernation)}
\scriptsize
\begin{tabulary}{1.0\textwidth}{C|CCC|CC|CC|CC}
ID & \multicolumn{3}{c|}{Hardware Configuration} &\multicolumn{2}{c|}{no-DBP} & \multicolumn{2}{c|}{DBP} & \multicolumn{2}{c}{MBS}\\
& MCU & Data Transceiver & Wake-Up & [$\mu$W] & Ratio & [$\mu$W] & Ratio  & [$\mu$W] & Ratio \\ \hline\hline
1 & Standby & LPM1 & none & \cellcolor[gray]{0.75}{5989} &  \cellcolor[gray]{0.75}1.0x & 3460 & 1.7x & 3453 & 1.7x\\ \hline
2 & Standby & LPM2 & none & 3532 &  1.7x & 759 & 7.9x & 749 & 8.0x\\ 
3 & Standby & LPM2+FF & none & 2988 & 2.0x & 759 & 7.9x & 749 & 8.0x\\ \hline
4 & Sleep & LPM2 & none & 3520 & 1.7x &  746 & 8.0x & 736 & 8.1x \\ 
5 & Sleep &  LPM2+FF & none &  2976 & 2.0x & 745 & 8.0x & 736 & 8.1x\\ \hline
6 & Sleep & LPM2 & US & 2346 & 2.6x & 15.4 & 388x & 5.4 & 1102x\\ 
7 & Sleep & LPM2+FF &  US & 1795 & 3.3x & 15.1 & 397x & 5.1 & 1175x\\
8 & Sleep & LPM2 & USa & - & - & 15.7 & 383x & 5.7 & 1060x\\ \hline
9 & Sleep & LPM2 & Radio & 1696 & 3.5x & 13.8 & 435x & 3.8 & 1591x\\
10 & Sleep & LPM2+FF & Radio & 1145 & 5.2x & 13.4 & 446x & 3.4 & 1748x\\ 
11 & Sleep & LPM2 & Radioa & 318 & {\bf 19x} & 12.9 & {\bf 463x} & 2.9 & {\bf 2052x}\\ \hline
12 & Hib. & LPM2 & US & - & - & 1664 & 3.6x & 57.3 & 104x\\ 
13 & Hib. & LPM2+FF & US & - & - & 1663 & 3.6x & 57.0 & 105x\\ 
14 & Hib. & LPM2 & USa & - & - & 1617 & 3.7x & 10.3 & 582x\\ \hline
15 & Hib. & LPM2 & Radio & - & - & 1663 & 3.6x & 56.9 & 105x\\ 
16 & Hib. & LPM2+FF & Radio & - & - & 1663 & 3.6x & 56.5 & 106x\\ 
17 & Hib. & LPM2 & Radioa & 10042 & 0.6x & 1617 & 3.7x & 8.9 & 676x\\ \hline
18 & \tiny{ML Hib.} & LPM2 & US & 2424 & 2.5x & - & - & 4.5 & 1324x\\ 
19 & \tiny{ML Hib.} & LPM2+FF & US & 1873 & 3.2x & - & - & 4.2 & 1431x\\ 
20 & \tiny{ML Hib.} & LPM2 & USa & - & - & - & - & 4.7 & 1275x\\ \hline
21 & \tiny{ML Hib.} & LPM2 & Radio & 1775 & 3.4x & - & - & 4.1 & 1444x\\ 
22 & \tiny{ML Hib.} & LPM2+FF & Radio & 1224 & 4.9x & - & - & 3.8 & 1572x\\ 
23 & \tiny{ML Hib.} & LPM2 & Radioa & 328 & 18x & - & - & 3.3 & 1837x\\ \hline
\end{tabulary}
\vspace{-6mm}
\label{tab:results1}
\end{center}
\end{table*}

\begin{table*}[htb]
\begin{center}
\caption{System-wide energy savings in the Intel-lab case study (light). The gray cells
  indicate the baseline for calculating the power consumption improvement ratio of all
  other configurations. (Hib.=Hibernation)}
\scriptsize
\begin{tabulary}{1.0\textwidth}{C|CCC|CC|CC|CC}
ID & \multicolumn{3}{c|}{Hardware Configuration} &\multicolumn{2}{c|}{no-DBP} & \multicolumn{2}{c|}{DBP} & \multicolumn{2}{c}{MBS}\\
& MCU & Data Transceiver & Wake-Up & [$\mu$W] & Ratio & [$\mu$W] & Ratio  & [$\mu$W] & Ratio \\ \hline\hline
1 & Standby & LPM1 & none & \cellcolor[gray]{0.75}{6323} &  \cellcolor[gray]{0.75}1.0x & 3529 & 1.8x & 3522 & 1.8x\\ \hline
2 & Standby & LPM2 & none & 3900 &  1.6x & 834 & 7.6x & 825 & 7.7x\\ 
3 & Standby & LPM2+FF & none & 2996 & 2.1x & 812 & 7.8x & 803 & 7.9x\\ \hline
4 & Sleep & LPM2 & none & 3888 & 1.6x &  821 & 7.7x & 812 & 7.8x \\ 
5 & Sleep &  LPM2+FF & none &  2984 & 2.1x & 799 & 7.9x & 790 & 8.0x\\ \hline
6 & Sleep & LPM2 & US & 2530 & 2.5x & 75.1 & 84x & 65.6 & 96x\\ 
7 & Sleep & LPM2+FF &  US & 1615 & 3.9x & 52.7 & 120x & 43.3 & 146x\\
8 & Sleep & LPM2 & USa & - & - & 30.3 & 208x & 20.9 & 303x\\ \hline
9 & Sleep & LPM2 & Radio & 2368 & 2.7x & 69.9 & 91x & 60.4 & 105x\\
10 & Sleep & LPM2+FF & Radio & 1453 & 4.3x & 47.6 & 133x & 38.1 & 166x\\ 
11 & Sleep & LPM2 & Radioa & 68.7 & {\bf 92x} & 13.8 & {\bf 457x} & 4.4 & {\bf 1442x}\\ \hline
12 & Hib. & LPM2 & US & - & - & 4755 & 1.3x & 3234 & 2.0x\\ 
13 & Hib. & LPM2+FF & US & - & - & 4733 & 1.3x & 3212 & 2.0x\\ 
14 & Hib. & LPM2 & USa & - & - & 1600 & 4.0x & 79.1 & 80x\\ \hline
15 & Hib. & LPM2 & Radio & - & - & 4752 & 1.3x & 3230 & 2.0x\\ 
16 & Hib. & LPM2+FF & Radio & - & - & 4729 & 1.3x & 3208 & 2.0x\\ 
17 & Hib. & LPM2 & Radioa & - & - & 1585 & 4.0x & 63.8 & 99x\\ \hline
18 & \tiny{ML Hib.} & LPM2 & US & 2648 & 2.4x & - & - & 67.5 & 94x\\ 
19 & \tiny{ML Hib.} & LPM2+FF & US & 1733 & 3.6x & - & - & 45.4 & 140x\\ 
20 & \tiny{ML Hib.} & LPM2 & USa & - & - & - & - & 20.0 & 316x\\ \hline
21 & \tiny{ML Hib.} & LPM2 & Radio & 2487 & 2.5x & - & - & 63.6 & 99x\\ 
22 & \tiny{ML Hib.} & LPM2+FF & Radio & 1572 & 4.0x & - & - & 41.3 & 153x\\ 
23 & \tiny{ML Hib.} & LPM2 & Radioa & 71.1 & 89x & - & - & 4.7 & 1325x\\ \hline
\end{tabulary}
\vspace{-6mm}
\label{tab:results2}
\end{center}
\end{table*}

\subsection{Comparison of Wake-up Receivers}
\begin{table}[tb]
\begin{center}
\caption{Benefits of using the the radio WuR over the US WuR.}
\scriptsize
\begin{tabular}{|c|c|c|c|} \hline
{\bf Case study} & {\bf \parbox{1.5cm}{Dataset}} &{\bf \parbox{3.9cm}{Avg. Network Data Rate}} & {\bf \parbox{2.7cm}{Power Savings}} \\ \hline
 & Light & 152 pkts/hour & 54.38\%   \\ 
\indoor & Humidity & 32 pkts/hour &  26.01\%  \\ 
 & Temperature & 26 pkts/hour & 22.66\%\\ \hline
\tunnel & Light & 13 pkts/hour & 14.42\%\\ \hline
\end{tabular}
\vspace{-6mm}
\label{tab:wur_vs_us}
\end{center}
\end{table}

System performance is sensitive to the WuRs and the network traffic. As highlighted by
Table~\ref{tab:wurscomparison}, the radio WuR offers three orders
of magnitude higher throughput than the US WuR and therefore can
send and receive the triggering signal more quickly, saving energy. For this
reason, the radio WuR \emph{always} achieves higher energy efficiency  than the US WuR. 
 
Table~\ref{tab:wur_vs_us} reports the percentage reduction in power consumption because of using the most efficient configuration of the radio WuR over that of the US WuR. For the denser \indoor deployment, the addressing modes of the both WuRs (USa and Radioa) are more efficient than the broadcast mode (US and Radio), which unnecessarily wakes up many neighbors.

Compared to USa (ID 8, \dbp), Radioa (ID 11, \dbp) reduces the average node power consumption by 54.38\%
for light, 26.01\% for humidity and 22.66\% for temperature, as shown in
Table~\ref{tab:wur_vs_us}.  The larger improvements represent the datasets generating more \dbp models. 
For example,  \indoor light is the most difficult dataset to be predicted by \dbp, generating $152$ models per hour compared to
only 32 and 26 packets per hour for humidity and temperature. Therefore, relatively larger number of \dbp models benefit from the efficient radio wake-up triggering, resulting in more power savings compared to the humidity and the temperature datasets. A better energy efficiency of USa than Radioa is also clearly reflected on the lifetime improvement observed for the humidity and temperature datasets.

For \tunnel, the improvement achieved by the radio WuR (ID 11, \dbp) over US WuR (ID 7, \dbp) is significant (14.42\%) but less than the one for the \indoor dataset. The efficiency advantage of the radio WuR over the US WuR is limited by two factors.  First, \dbp suppresses 99.74\% periodic transmissions, generating \emph{only} 13 packets per hour in the \tunnel network on average. Second, most traffic comes from the nodes exposed to direct sunlight at the entrance of tunnel (e.g., 1-3 and 21-23). Because the gateway is deployed at the entrance, the direct links between these nodes and the sink can exploit the efficient WuRs over only a single hop in most cases and that also for ultra-low traffic. Both factors limit the advantage of the radio WuR over the US WuR to 14.42\% in this particular scenario. However, it is worth-mentioning that the WuRs achieve a remarkable two orders of magnitude reduction (397x in ID 7, 463x in ID 11) in power consumption compared to the base configuration by cutting the costs of idle listening and overhearing.

%

To sum up, we evaluated our architecture with two different WuRs, highlighting better energy efficiency of the radio WuR over the US WuR. In addition, an omni-directional nature of the radio WuR brings other advantages in that a data collection tree can be constructed on-demand and adapted to dynamic network conditions. By extending our evaluation of the US WuR~\cite{sies14} to the radio WuR, we show a considerable improvement  in our system performance. Thanks to the modular nature of the VirtualSense mote, integrating new WuRs is straight-forward.

 

%
%
%
%

\subsection{Performance of \dbs}

\begin{figure}[!t]
\vspace{-4mm}
\centering
\mbox{
 \subfigure[US.]{
     \includegraphics[scale=0.45, Trim=2cm 8cm 15cm 2cm, clip=true]{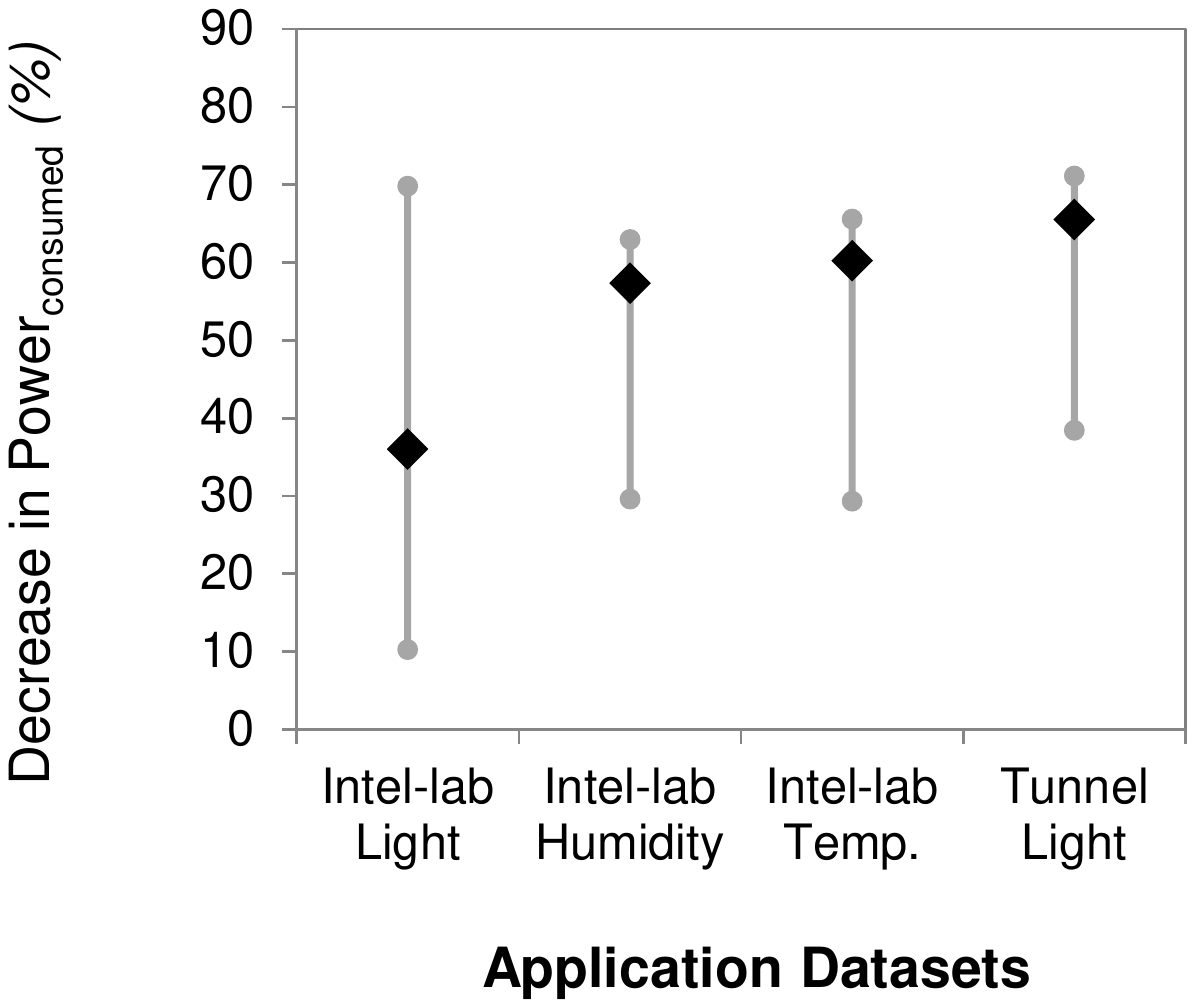}     
     \label{fig:sensinglayer_improvements_us}
   }
 \subfigure[Radio.]{
    \includegraphics[scale=0.45, Trim=4cm 8cm 15cm 2cm, clip=true]{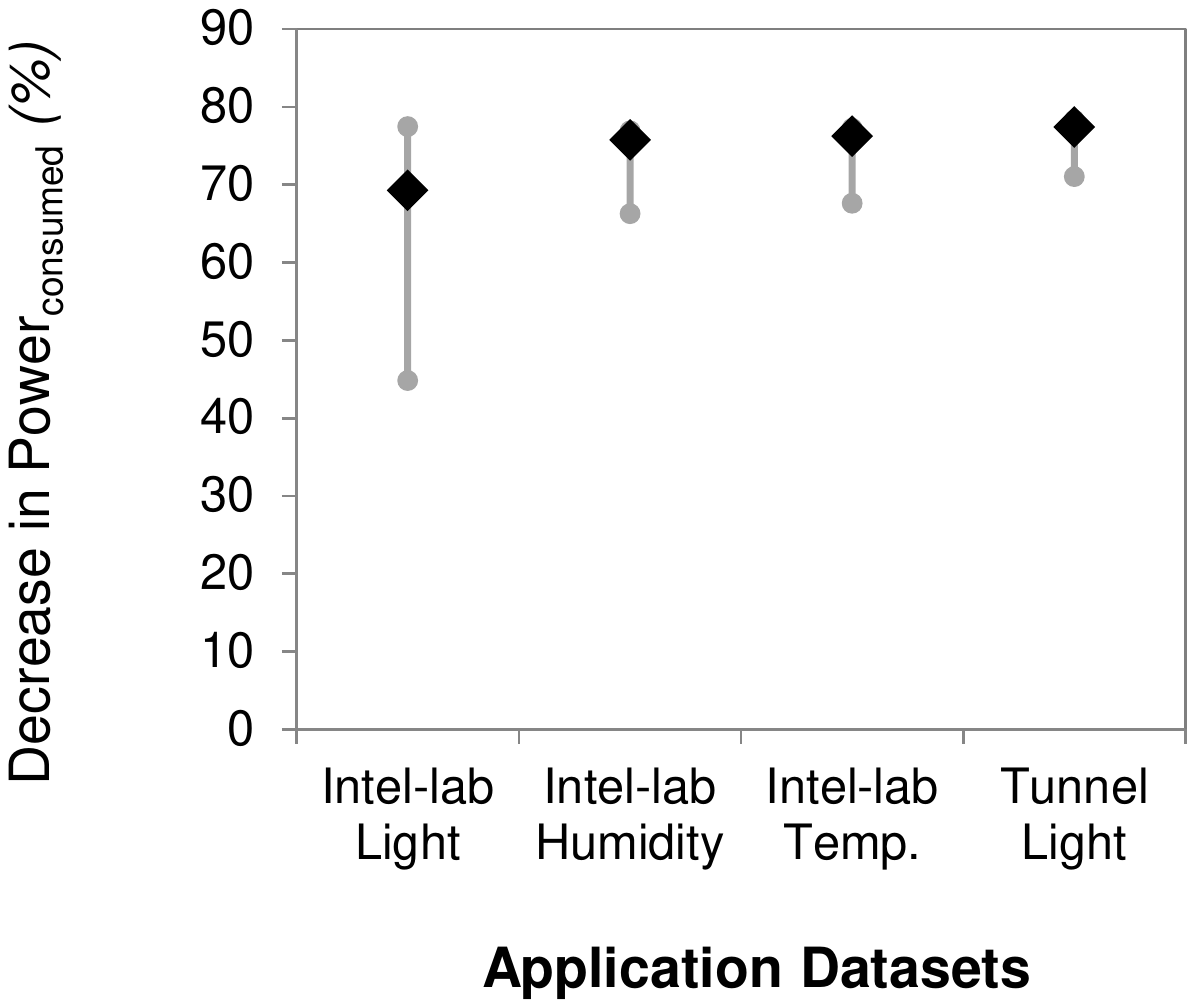}
     \label{fig:sensinglayer_improvements_radio}
   }
}
\vspace{-4mm}
\caption{Average, minimum and maximum reduction in power consumption enabled by the separate sensing layer for the network nodes  \whereis{the first two sheets of /img/graph-data/SensingLayer\_improvements.xlsx}}
\label{fig:sensinglayer_improvements}
\end{figure}

Thanks to the combination of \dbp and WuRs, we greatly reduce the communication cost, leaving sampling cost to dominate the node consumption. Therefore, we now turn our attention to the ability of \dbs in reducing the sampling cost. With \dbs, a hardware peripheral can sense and process the samples without involving the power-hungry MCU of the mote as long as \dbp needs not to update the model. This enables significant savings, as explained next. 

Figure~\ref{fig:sensinglayer_improvements} shows the percentage, minimum and maximum power savings enabled by \dbs for the network nodes. We notice that the average node power consumption improves remarkably from 35\% to 76\%, irrespective of the WuR technology.  Moreover, the minimum improvement for a single node is significant, above 45\% when a radio WuR is used. The lower energy efficiency of US WuR and a higher \dbp traffic rate for \indoor datasets shift the overall cost from sampling to communication, thus slightly dropping the percentage power savings enabled by our sampling peripheral. Overall, when \dbs is combined with other techniques from our architecture, the average power consumption drops from milliwatts to a few microwatts. Specifically, Tables~\ref{tab:results1}-\ref{tab:results2} highlight that the average power consumption is reduced merely to $2.9\mu_W$ and $4.4\mu_W$ (ID 11, MBS) for \tunnel and \indoor respectively.



%

%
%
%

\subsection{Energetic Sustainability}
\begin{figure}[!t]
\vspace{-4mm}
\centering
\includegraphics[width=0.8\columnwidth, Trim=2.1cm 2cm 2cm 2cm, clip=true]{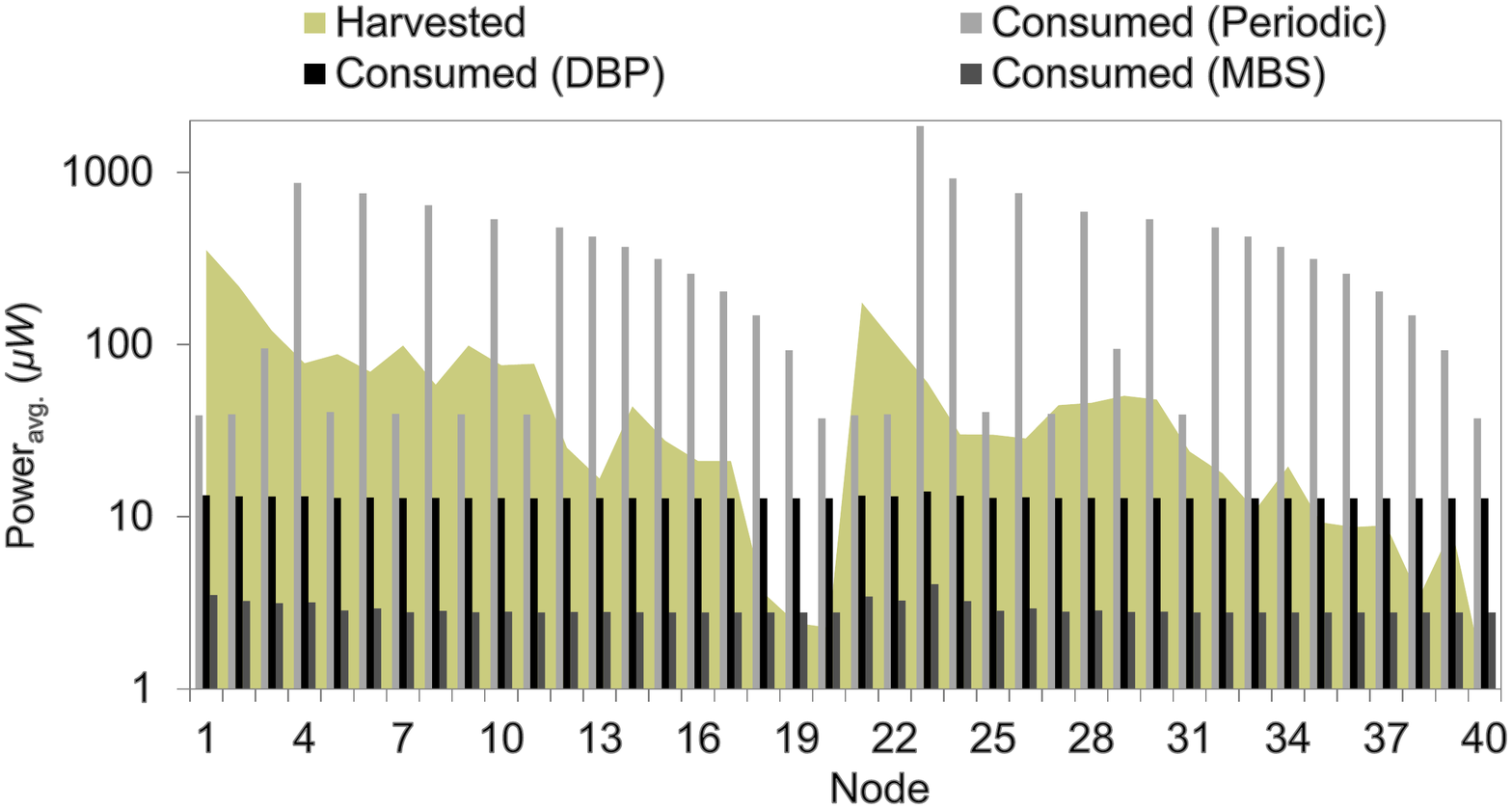} 
\includegraphics[width=0.8\columnwidth, Trim=2.2cm 1cm 2cm 6.5cm, clip=true]{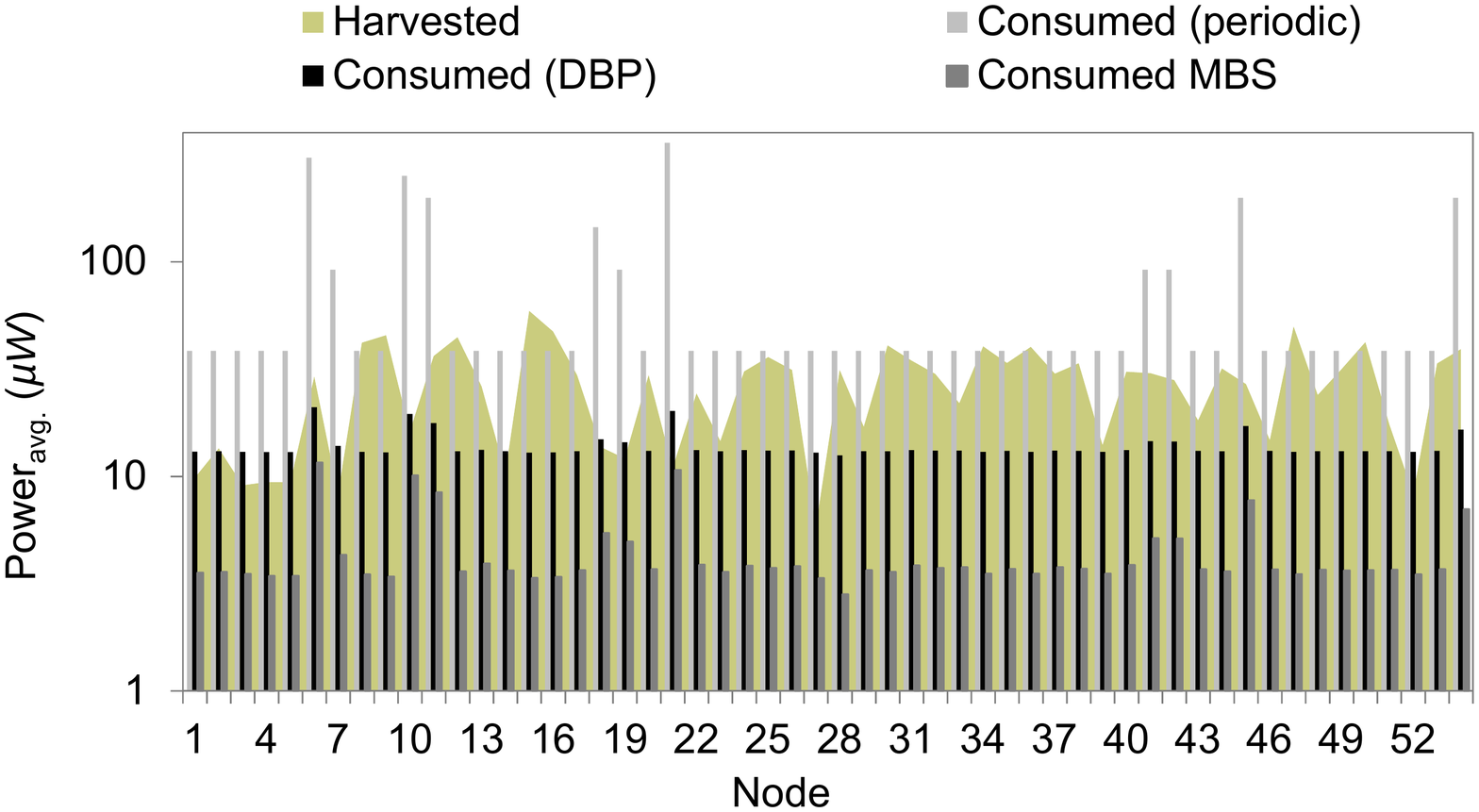} 

\vspace{-6mm}
\caption{\memo{Comparison of energy consumed and harvested, plotted on a logarithmic
  scale for \tunnel (up) and \indoor (down)}.  \whereis{/img/graph-data/HarvVsCons\_Tunnel\_WUR.xlsx} }
\vspace{-2mm}
\label{fig:consVsHarvtunnel}
\end{figure}

Next we return to our initial motivation: namely the creation of an
energy-neutral system with a reasonable, indoor energy harvester. Referring to
Table~\ref{tab:indoorharvest}, we note that the power density of an indoor
harvester is typically less than a few microwatts/cm$^2$. Therefore, to keep
the size of the harvester similar to the size of a node, the overall power
consumption should be on the same order. 

We first consider this specifically for the tunnel dataset. 
Figure~\ref{fig:consVsHarvtunnel} shows the power consumed in \tunnel in three
different software/hardware configurations against the energy harvested by a
single photovoltaic cell (namely a Panasonic AM-1816~\cite{am1816}). When
nodes do not use \dbp, all the sensed values are sent to the sink and most
nodes consume more than they can harvest due to large forwarding overhead,
making them not energetically sustainable. Using \dbp without using \dbs
allows most nodes to consume less than they harvest. Nevertheless, several
nodes deep in the tunnel harvest so little energy that they would need a large
number of solar panels to be energetically sustainable. Overall, we would
require 72 photovoltaic cells to sustain 40 nodes. Instead, if we consider 
the hardware sensing layer, \dbs, almost all nodes can
sustain their operation with only a single photovoltaic cell, reducing the
requirement from 72 to only 44 cells, a significant savings.

\memo{We now turn our attention to \indoor network. Compared to \tunnel, it is characterized by a shallow data collection topology with a maximum diameter of only four hops. As most nodes are directly connected to the gateway (see Figure~\ref{fig:topology_intel})  and only a few nodes relay any traffic, their overall power consumption is far less than the \tunnel nodes. Furthermore, an analysis of light traces collected from Intel Research Lab. suggests that these nodes are exposed to a higher illuminance. It means that smaller photovoltaic cells can be used to achieve energy neutrality, save costs, and miniaturize the sensor node design. Figure~\ref{fig:consVsHarvtunnel} shows the energy harvested by the same but ten times smaller photovoltaic harvester than the one used for \tunnel.  When the nodes report periodic data, all forwarding nodes as well as many non-forwarding nodes consume more power than they can harvest. By decreasing the overall network traffic substantially, \dbp reduces the consumed power below the harvested power for most but not all the nodes. \dbs, instead, makes all the nodes energy neutral, enabling cost-efficient perpetual data collection.}


\section{Conclusion}
\label{sec:conclusions}

Obtaining energy neutral operation for WSNs is a challenging task, especially in
indoor settings, due to the sampling and communication tasks that require nodes
to exit their extreme low power states. This paper introduced the concept of
model-based sensing and demonstrated its effectiveness to reduce power
consumption when combined with several other state-of-the-art techniques.
Specifically, advanced dynamic power management is adopted at each node,
prediction-based data collection is used to reduce data traffic by avoiding
the transmissions of data fitting the current model, wake-up receivers are
used to avoid idle listening in asynchronous communication, and data storage
and processing capabilities are granted to the hardware sampling layer to
allow it to discard useless samples without waking up the main MCU.

Experimental results using data conducted in two real-world WSN case studies
show that the synergistic application of the different techniques offers power
saving of up to three orders of magnitude, representing a significant
improvement with respect to the state of the art. Most important, the proposed
approach brings the average power consumption of the WSNs used in our case
studies within the power budget of indoor photovoltaic harvesters, thus
achieving energy neutral operation. This study concretely
demonstrates that combining hardware and software techniques is key to energy
neutrality for energy harvesting WSNs deployed in indoor conditions.

\section*{Acknowledgements}
This work was partially funded by the European Institute of Innovation \& Technology (EIT ICT Labs---Activity 15171.).

\section*{References}
\bibliographystyle{elsarticle-num}
\memo{
\bibliography{paper}
}
\end{document}